\pdfoutput=1
\documentclass[%
 reprint,
superscriptaddress,
 amsmath,amssymb,
 aps,
pra,
]{revtex4-2}

\usepackage{float}
\usepackage{amssymb} 
\usepackage{subfigure}
\usepackage{xcolor}
\usepackage{graphicx}
\usepackage{dcolumn}
\usepackage{bm}
\usepackage{physics}
\usepackage{siunitx}
\usepackage{ulem}
\usepackage{mathrsfs}
\usepackage[mathlines]{lineno}  

\renewcommand{\vec}[1]{\mathbf{#1}}

\newcommand{\appropto}{\mathrel{\vcenter{
  \offinterlineskip\halign{\hfil$##$\cr
    \propto\cr\noalign{\kern2pt}\sim\cr\noalign{\kern-2pt}}}}}

\begin{document}

\title{Computational aberration-retrieval with entangled photons}

\author{Baptiste Courme} 
\thanks{These authors contributed equally}
    \affiliation{Sorbonne Université, CNRS, Institut des NanoSciences de Paris, INSP, F-75005 Paris, France}
    \affiliation{Laboratoire Kastler Brossel, ENS-Université PSL, CNRS, Sorbonne Université, Collège de France, 24 rue Lhomond, 75005 Paris, France}

\author{YoonSeok Baek}
\thanks{These authors contributed equally}
    \affiliation{Laboratoire Kastler Brossel, ENS-Université PSL, CNRS, Sorbonne Université, Collège de France, 24 rue Lhomond, 75005 Paris, France}

\author{Hugo Defienne}
    \email[Corresponding author: ]{hugo.defienne@insp.upmc.fr}
    \affiliation{Sorbonne Université, CNRS, Institut des NanoSciences de Paris, INSP, F-75005 Paris, France}

\begin{abstract}
Adaptive optics (AO) is central to high-resolution microscopy, yet conventional implementations often require repeated correction-and-measurement cycles whose number increases with aberration complexity. 
This measurement burden can become prohibitive when the available temporal window for correction is short or when photon budgets are constrained. 
Here, we demonstrate an AO strategy that exploits the spatial correlations of entangled photon pairs to retrieve aberrations directly. 
By using the second-order correlation function as an aberration-sensitive metric, the required correction can be determined with only a few measurements, without relying on image-based feedback or iterative optimization. 
This correlation-based approach establishes an efficient framework for AO leveraging quantum light, offering a new route toward quantum-enhanced microscopy in regimes where conventional correction strategies become challenging.
\end{abstract}

\maketitle

\section{Introduction}

In microscopy, non-classical optical sources can be used to surpass standard imaging limits, offering enhanced spatial resolution~\cite{he_quantum_2023}, higher signal-to-noise ratios~\cite{casacio_quantum-enhanced_2021} and improved phase sensitivity~\cite{camphausen_quantum-enhanced_2021}. 
They also enable novel modalities such as ghost imaging~\cite{davenport_quantum_2024,eshun_3d_2025,ryan_infrared_2024}, imaging with undetected photons~\cite{kviatkovsky_microscopy_2020,pearce_practical_2023,vanselow_mid-infrared_2019} and quantum holography~\cite{defienne_polarization_2021,zhang_quantum_2024}. 
Building on this, Cameron \textit{et al.}~\cite{cameron_adaptive_2024} recently demonstrated a sensorless adaptive optics (AO) approach for microscopy driven by the spatial correlations of entangled photon pairs. Unlike conventional sensorless AO, this method corrects aberrations without requiring image-based metrics or guide stars~\cite{zhang_adaptive_2023, ji_adaptive_2017,booth_adaptive_2007}. However, like its classical counterparts, it remains fundamentally limited by its reliance on iterative hardware optimization.

Indeed, optimizing physical correction elements - a deformable mirror (DM) or spatial light modulator (SLM) - in sensorless AO requires numerous sequential measurements, a number that scales directly with aberration complexity. 
This measurement burden becomes a major constraint when the available imaging or correction window is limited.
Although several recent approaches aim to reduce the number of correction steps, they introduce significant trade-offs. Machine-learning methods~\cite{hu_universal_2023,zhang_single_2025}, for example, strongly depend on a pre-training phase requiring datasets carefully tailored to the specific optical system (e.g. numerical aperture and wavelength) and sample type. 
Similarly, phase-diversity-based AO algorithms~\cite{johnson_phase-diversity-based_2024} provide a fast alternative, but they rely on solving an inherently ill-posed inverse problem: the joint estimation of sample morphology and system aberrations. As a result, these methods are highly susceptible to convergence toward local minima, especially in highly scattering or low-contrast regimes.

Here, we demonstrate a non-iterative phase-retrieval approach that implements AO using only a few measurements, no more than five, of the spatial correlations between entangled photon pairs.
Our method exploits the fact that the second-order intensity correlation of photon pairs provides a direct, object-independent signature of the system’s point spread function (PSF). 
By decoupling the system aberrations from the sample morphology, the approach enables rapid algorithmic convergence without requiring prior models or system-specific pre-training. 
We experimentally validate the technique by imaging biological samples in a full-field transmission configuration, successfully correcting aberrations of varying complexity.

\section{Principle}

\begin{figure*}
    \includegraphics[width=0.75\linewidth]{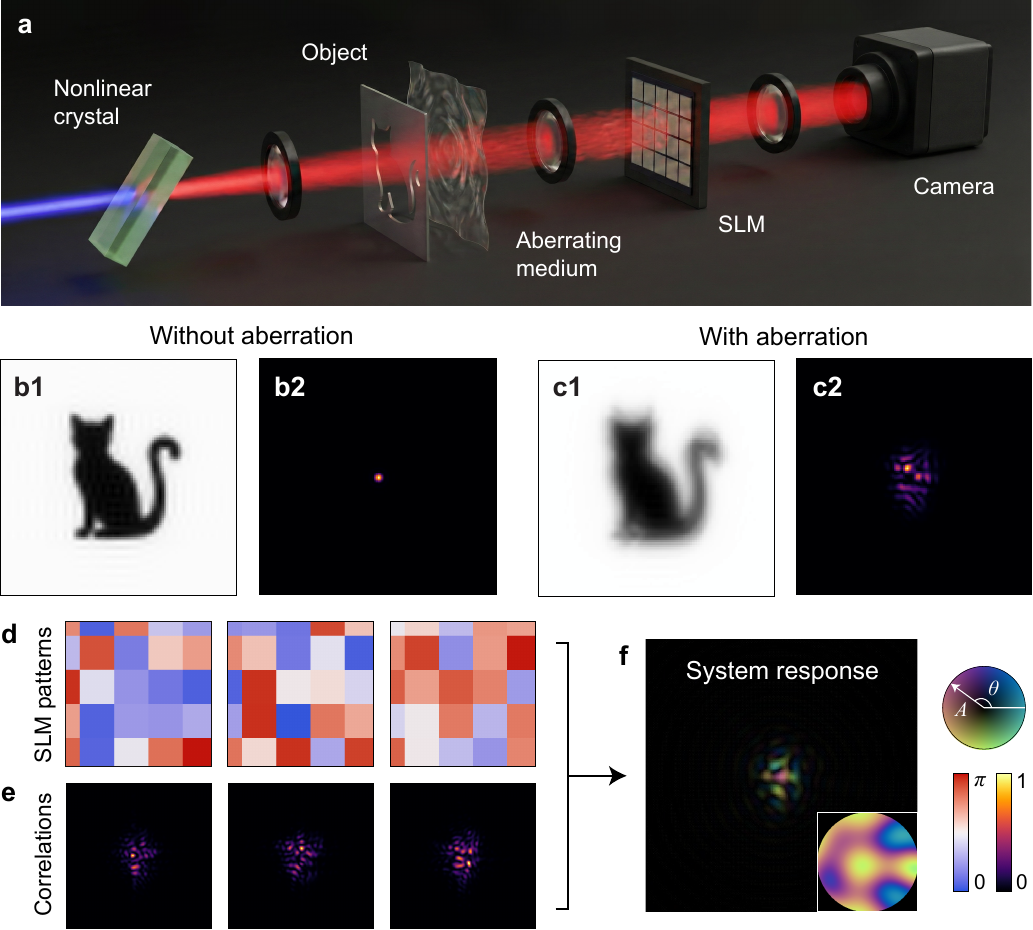}
  \caption{\textbf{Principle and simulations.} \textbf{(a)} An object is illuminated by a source of entangled photon pairs exhibiting strong spatial correlations. In the example considered here, the photons are strongly anti-correlated in the object plane i.e. when a signal photon is detected at position $\vec{r_s}$, the associated idler photon is detected around position $\vec{r_i} \approx -\vec{r_s}$. Such illumination can be generated, for example, by placing the object in the Fourier plane of a thin nonlinear crystal illuminated by a large-diameter pump laser. In this configuration, photon pairs are produced via spontaneous parametric down-conversion (SPDC), and the strong spatial anti-correlations are ensured by momentum conservation. The object is imaged onto a camera via a two-lens system, with a spatial light modulator (SLM) positioned between the lenses in the Fourier plane (pupil plane). 
  \textbf{(b)} In the absence of aberrations, the simulated intensity image yields a sharp representation of the object (b1), while the sum-coordinate projection $C^+$ exhibits a sharp central peak (b2), as predicted by Eq.~\ref{eq1}. 
  \textbf{(c)} In the presence of complex aberrations in the imaging path, the intensity image becomes severely blurred (c1), whereas the sum-coordinate projection degrades into a speckle pattern (c2). \textbf{(d)} To retrieve the phase associated with the system aberrations, distinct random phase patterns are displayed on the SLM. In this simulation, for example, we employ three such patterns. 
  \textbf{(e)} These patterns introduce known optical perturbations that correspondingly modulate the $C^+$ correlation images. \textbf{(f)} A phase-retrieval algorithm processes these modulated images to reconstruct the complex system response. The main panel displays the reconstructed coherent point spread function (PSF), while the inset reveals the associated phase profile in the pupil plane. The color wheel indicates the amplitude (brightness) and phase (hue). Details about the algorithm, simulations and experimental design are in Methods.}
  \label{Figure1}
\end{figure*}

The imaging system under consideration is illustrated in Fig.~\ref{Figure1}. 
In such a configuration, both a conventional intensity image ($I$) and the second-order intensity correlation function ($G^{(2)}(\vec{r_s},\vec{r_i})$) are retrieved from the same measurement sequence. The former is generated by accumulating photon counts at each pixel, while the latter is extracted via multi-pixel photon coincidence counting~\cite{defienne_general_2018,ndagano_imaging_2020}, where $\vec{r_s}=(x_s,y_s)$ and $\vec{r_i}=(x_i,y_i)$ are the transverse spatial positions of the signal and idler photons.
Here, we perform aberration correction exclusively using $G^{(2)}$. 
Indeed, as previously demonstrated in Ref.~\cite{cameron_adaptive_2024}, for strongly anti-correlated photon pairs (i.e. generated using a thin nonlinear crystal and a wide pump beam~\cite{abouraddy_entangled-photon_2002}), assuming shift invariance is preserved and the aberrations are not too complex, the projection of $G^{(2)}$ along the sum-coordinate transverse spatial axis, denoted $C^+$, is entirely independent of the spatial structure of the object. It can be expressed as
\begin{equation}
C^+(\vec{r^{+}}) \propto|H(\vec{r^{+}})|^2,
\label{eq1}
\end{equation}
where $H=h*h$ is the effective transfer function, $*$ denotes the convolution operator, $h$ is the coherent point spread function (PSF) of the imaging system linking the object plane to the camera plane. 
In practice, $C^+$ takes the form of a bi-dimensional image, with $\vec{r^{+}} = (x^+,y^+) = (x_s+x_i,y_s+y_i)$ being the sum of the signal $(x_s,y_s)$ and idler photons $(x_i,y_i)$ coordinates. 
Upon introducing aberrations, the sharp peak initially present at the center of $C^{+}_n$ (Fig.~\ref{Figure1}b2) becomes highly distorted (Fig.~\ref{Figure1}c2). However, because this measurement lacks phase information, both the effective transfer function $H$ and, consequently, the underlying coherent PSF $h$ remain undetermined.

To resolve this ambiguity, we introduce controlled modulation to the system response. 
Specifically, we apply a set of $N$ known perturbations $\{d_n\}_{n \in [1,N]}$ using the spatial light modulator (SLM), such that the modified system response becomes $h_n=h*d_n$ (Fig.~\ref{Figure1}d).
Each perturbation induces a predictable change in the effective transfer function $H$, where the $n$-th modulated sum-coordinate projection satisfies
\begin{equation}
    C^{+}_n(\vec{r^{+}})\propto|H*D_n(\vec{r^{+}})|^2,
    \label{eq.sum_projection_modulated}
\end{equation}
with $D_n=d_n*d_n$.
Unlike the unperturbed case, these measurements mix the unknown $H$ with known functions, encoding phase information into measurable correlation variations (Fig.~\ref{Figure1}e).
Based on Eq. (\ref{eq.sum_projection_modulated}), we can formulate a phase retrieval problem that reconstructs $H$ from a set of magnitude measurements $C^{+}_n$ subject to known multiplicative constraints at the Fourier plane (see Methods).
Once the effective transfer function $H$ is retrieved, the system impulse response is obtained as
\begin{equation}
    h(\vec{r})=\mathcal{F}^{-1}\left[\left|\mathcal{F}{(H)}\right|e^{i\theta(\vec{r})}\right],
\end{equation}
where $\mathcal{F}$ is the Fourier transform operator and $\theta(\vec{r})=\frac{1}{2}\textnormal{unwrap}\{\textnormal{arg}[\mathcal{F}{(H)}]\}$.
Similar to conventional phase-retrieval methods based on random or coded intensity measurements \cite{candes2015phase, wang2018phase, baek2020speckle}, which generally require an effective oversampling of about threefold for stable reconstruction, our method uses a minimum of three perturbations. Using more perturbations further improves robustness to noise. In this work, we found that 3–5 perturbations were sufficient across all simulations and experiments.

\section{Experimental test of aberration retrieval}

\begin{figure*}
    \includegraphics[width=0.8\linewidth]{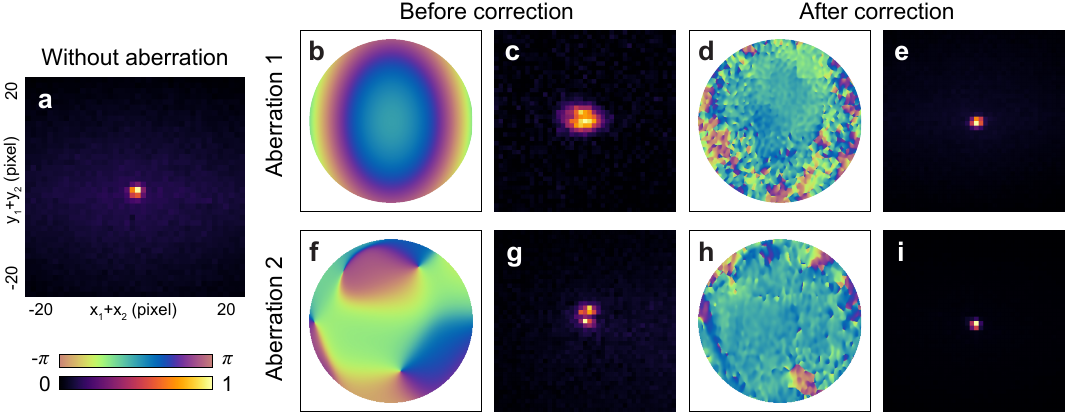}
  \caption{\textbf{Experimental results of aberration correction.} \textbf{(a)} Experimentally measured $C^+$ correlation image in the absence of aberrations. \textbf{(b, f)} Phase masks programmed onto the SLM to artificially induce aberrations in the imaging system. The masks are constructed by superimposing Zernike modes corresponding to defocus and astigmatism for Aberration 1 \textbf{(b)}, and by using the phase of complex Gaussian speckle field with a speckle grain size of 150 SLM pixels for Aberration 2 \textbf{(f)}. \textbf{(c, g)} Corresponding distorted $C^+$ correlation images measured prior to correction. \textbf{(d, h)} Phase masks displayed on the SLM resulting from the subtraction of the algorithm-retrieved phase from the initially programmed aberration masks. \textbf{(e, i)} Corresponding $C^+$ correlation images measured after correction, demonstrating the successful restoration of the sharp central peak. } 
  \label{Figure2}
\end{figure*}

For the experimental demonstration, we use the optical configuration shown in Fig.~\ref{Figure1}. Entangled photon pairs are generated in a 0.5-mm-thick type-I $\beta$-barium borate (BBO) nonlinear crystal, pumped by a continuous-wave laser at 405 nm with a beam diameter of approximately 1 mm. 
A long-pass filter with a 750 nm cut-on wavelength is positioned immediately after the crystal to reject the pump photons while transmitting the frequency-degenerate photon pairs centered around 810 nm. 
The first lens, which Fourier-images the crystal onto the object plane, has a focal length of 300 mm. The subsequent two lenses form the imaging system and provide a magnification of $M=5$ (see Methods for full details). 
Photon pairs are detected using a $32 \times 64$-pixel single-photon avalanche diode (SPAD) array camera, which enables the direct measurement of the correlation function $G^{(2)}$ and its sum-coordinate projection $C^+$~\cite{ndagano_imaging_2020}. An $810 \pm 10$ nm band-pass filter is placed directly in front of the camera.

The aberration-retrieval algorithm is initially tested without an object by artificially inducing aberrations in the imaging path via the SLM. 
In the absence of aberrations, the $C^+$ correlation image (Fig.~\ref{Figure2}a) exhibits a sharp central peak, which is a direct signature of the strong spatial anti-correlations between the photon pairs. 
In the presence of aberrations, however, these correlations degrade severely. 
For example, Figs.~\ref{Figure2}c and g show distorted correlation images resulting from two distinct aberration profiles programmed onto the SLM (Figs.~\ref{Figure2}b and f). 
To correct these aberrations, we implement our aforementioned aberration-retrieval approach. 
In each scenario, five known phase masks, with a superpixel size of 80 × 80 SLM pixels and random phase values between 0 and $\pi$, were superimposed onto the existing aberration mask on the SLM, and the corresponding correlation images are recorded. 
These measurement sets allow us to estimate the system's coherent PSF, $h$. 
By applying a Fourier transform, we then retrieve the associated phase profile in the pupil plane. 
Finally, subtracting this retrieved phase from the initial aberration mask (Figs.~\ref{Figure2}d and h) successfully restores the sharp correlation peak in the output images (Figs.~\ref{Figure2}e and i), directly demonstrating the high efficacy of the correction.

\section{Application to adaptive optics}

\begin{figure*}
    \centering
    \includegraphics[width=0.7\linewidth]{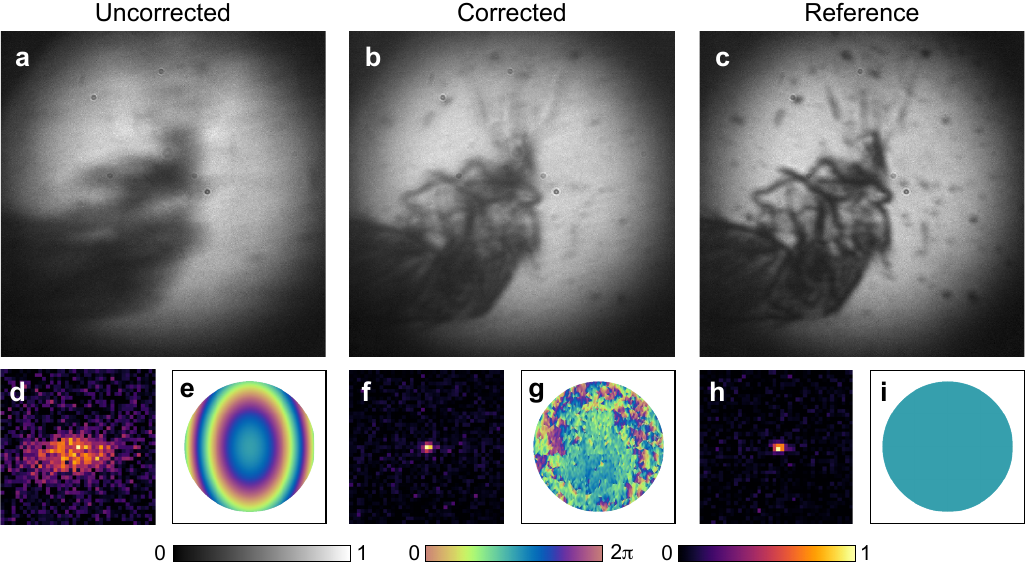}
  \caption{\textbf{Experimental results of adaptive optics.} In the presence of aberrations, the intensity image \textbf{(a)} becomes blurred, and the $C^+$ correlation image \textbf{(d)} is distorted. These aberrations are emulated by displaying a phase pattern \textbf{(e)} on the SLM, constructed by superimposing Zernike polynomials for defocus and astigmatism. After applying our approach, the retrieved phase correction is superimposed onto the initial aberration pattern on the SLM, resulting in a nearly flat effective phase profile \textbf{(g)}. Consequently, the corrected intensity image \textbf{(b)} is sharp, and the $C^+$ correlation image \textbf{(f)} exhibits a restored sharp central peak. For comparison, the reference intensity image \textbf{(c)} and $C^+$ correlation image \textbf{(h)} are recorded in the absence of aberrations, corresponding to a perfectly flat phase mask on the SLM \textbf{(i)}.}
    \label{Figure3}
\end{figure*}

To evaluate the potential of our approach under realistic imaging conditions, we insert a target object - specifically, a portion of an insect body mounted on a microscope slide - into the imaging path. 
Conventional intensity images are acquired using an electron-multiplying charge-coupled device (EMCCD) camera, which is positioned in a conjugate plane to the SPAD array and accessed via a flip mirror.
In the presence of optical aberrations (artificially induced via the SLM, as shown in Fig.~\ref{Figure3}e), the resulting intensity image of the object becomes severely blurred (Fig.~\ref{Figure3}a). 
The corresponding $C^+$ correlation image, measured simultaneously with the SPAD camera (Fig.~\ref{Figure3}h), exhibits a strongly distorted correlation peak consistent with the orientation of the applied aberration mask. Notably, while the presence of the object does not alter the underlying spatial profile of $C^+$, it absorbs a fraction of the photon flux at this wavelength, inherently leading to a reduction in the SNR.

After executing the aberration-retrieval procedure using five known random phase masks, the retrieved phase profile is subtracted from the initial aberration mask on the SLM (Fig.~\ref{Figure3}g). 
The resulting corrected phase mask is nearly flat over a central region spanning approximately 500 SLM pixels, which translates to a successful restoration of the insect's intensity image (Fig.~\ref{Figure3}b). 
In this case, the correction substantially restores the main spatial features of the object, although the recovery of the finest details remains limited. This behavior arises because the retrieved phase correction is most reliable over the central region of the effective pupil, while information from the outer pupil contributes less strongly. In our configuration, the SLM is conjugate to the nonlinear crystal generating the two-photon state, so the incident spatial intensity profile at the SLM is approximately Gaussian. As a result, the two-photon flux is highest near the center of the pupil, where the phase is retrieved and corrected most effectively, and decreases toward the periphery, where the correction becomes less complete.

\section*{Conclusion}

In conclusion, we demonstrate a computational algorithm that measures optical aberrations by exploiting the spatial correlations of entangled photon pairs. 
This approach succeeds because these correlations yield a direct, object-independent signature of the system's PSF - a critical advantage over sensorless wavefront-sensing methods. 
By drastically reducing the number of measurements required compared to standard iterative AO, our method shows strong potential for improving real-world and quantum-enhanced microscopy in dynamic environments, including in vivo imaging.

In its present form, however, our approach faces two major applicability constraints. 
First, it relies on the presence of strong spatial correlations between the photon pairs at the object plane. 
In our demonstration, this condition is met by directly illuminating the target with the incident two-photon state. In realistic scenarios, however, the target is typically embedded within a complex environment. Consequently, the illumination beam itself will also propagate through aberrations, weakening the spatial correlations at the object plane and, in turn, degrading the efficiency of our approach. 
To circumvent this issue, future implementations could isolate the ballistic illumination photons using time- or spatial-gating techniques~\cite{badon_smart_2016,kang_imaging_2015}, or incorporate a second SLM into the illumination or the pump path to pre-compensate for these incoming aberrations~\cite{lib_real-time_2020}.

Second, while our approach drastically reduces the number of measurements compared to iterative methods, this advantage is currently offset by long individual acquisition times. With our present SPDC source and SPAD array, achieving a sufficient signal-to-noise ratio for a single correlation image requires over $10^7$ frames i.e. several hours of acquisition, precluding real-time application. 
However, state-of-the-art time-stamping cameras have already compressed acquisition times to a few seconds under similar conditions~\cite{nomerotski_imaging_2019}.
Driven by ongoing advancements in high-brightness quantum sources and next-generation single-photon detectors~\cite{hogenbirk_intensified_2026,pitsch_toward_2025}, we expect these times to reach the millisecond regime in the coming years, ultimately establishing our approach as a practical, real-time adaptive optics solution for advanced \textit{in vivo} microscopy.

\section*{Methods}

\subsection{Iterative algorithm}
The algorithm iteratively finds the complex transfer function $H$ from the series of modulated sum projection of $G^{(2)}_n$.
It is based on the fact that the perturbation $d_n$ makes the system response in the Fourier plane as $\mathcal{F}[h_n]=\tilde{H}\tilde{D}_n$, where $\tilde{H}=\mathcal{F}(H)$ and $\tilde{D}_n=\mathcal{F}(D_n)$.
This allows us to interpret the sum projection in Eq. (\ref{eq.sum_projection_modulated}) as the Fourier magnitude information,
\begin{equation}
    C^{+}_n(\vec{\delta r^{+}})\propto|\mathcal{F}^{-1}[\tilde{H}\tilde{D}_n]|^2(\vec{\delta r^{+}}).
\end{equation}
The iteration begins by initializing the estimate $\tilde{H}_{est}^{(0)}$ with random complex values.
During each iteration $i$, the algorithm refines the estimate using an auxiliary variable $\tilde{\xi}_n^{(i)}=\tilde{H}_{est}^{(i)}\tilde{D}_n$.
The variable is updated by enforcing the Fourier domain constraint from the measured sum projection:
\begin{equation}
    \xi_n^{(i)}={\frac{\sqrt{C^{+}_n}}{|\xi_n^{(i-1)}|}\xi_n^{(i-1)}},
\end{equation}
where $\mathcal{F}[\xi^{(i)}] = \tilde{\xi}^{(i)}$.
Then the estimate of $\tilde{H}$ is updated
by compensating the known perturbation from $\xi_n$:
\begin{equation}
    \tilde{H}_{est}^{(i)}={\frac{\tilde{D}_n^*}{|\tilde{D}_n|^2}\tilde{\xi}_n^{(i)}}.
\end{equation}
Following these steps for different perturbations repeatedly, $\tilde{H}_{est}$ converges to the transfer function consistent with the measured sum projections.

While the iterative process above is sufficient in principle, experimental second-order correlation measurements are sensitive to detector noise and mechanical instability.
To improve robustness against these incoherent noise contributions, we introduce an additional step that models the measured sum projection as an incoherent sum of multiple transfer function estimates, $H_k$:
\begin{equation}
    C^{+}_n(\vec{\delta r^{+}})\propto\sum_k|H_k*D_n|^2(\vec{\delta r^{+}}).
\end{equation}
In an ideal, noise-free scenario, these estimates are identical. But, under noisy conditions, this multi-estimate allows the algorithm to isolate noise to a certain degree.
Each estimate is updated using individual auxiliary variables, $\tilde{\xi}_{k,n}^{(i)}=\tilde{H}_{est,k}^{(i)}\tilde{D}_n$
and the Fourier magnitude projection as
\begin{equation}
    \xi_{k,n}^{(i)}={\sqrt{\frac{C^{+}_n}{\sum_k |\xi_{k,n}^{(i-1)}|^2}}\xi_{k,n}^{(i-1)}},
\end{equation}
followed by
\begin{equation}
    \tilde{H}_{est,k}^{(i)}={\frac{\tilde{D}_n^*}{|\tilde{D}_n|^2}\tilde{\xi}_{k,n}^{(i)}}.
\end{equation}
This update is performed across all perturbations, multiple times.
Finally, to extract the coherent signal from the results, we apply singular value decomposition to $\bm{\tilde{H}}$ where each column corresponds to an individual estimate $\tilde{H}_{est,k}$:
\begin{equation}
 \bm{\tilde{H}}=\sum_{i}{\sigma_i}\vec{u}_{i}\vec{v}_{i}^{\dagger}.
\end{equation}
By filtering out the lower-order components associated with noise, the singular vector corresponding to the largest singular value, $\vec{u}_1$, becomes the final estimate of the system’s complex transfer function $\tilde{H}$.

\subsection{Experimental details}

The pump laser is a collimated continuous-wave laser operating at 405 nm (Coherent OBIS-LX), delivering an output power of 100 mW with a beam diameter of $0.8 \pm 0.1$ mm. Photon pairs are generated in a $0.5 \times 5 \times 5$ mm BBO crystal (Newlight Photonics), cut for type-I spontaneous parametric down-conversion (SPDC) at 405 nm with a half-opening angle of $3^\circ$. The crystal is slightly tilted about its horizontal axis to achieve near-collinear phase matching of the generated photons, causing the SPDC emission ring to collapse into a disk. 
Residual pump light is removed using a 650 nm long-pass filter, followed by a band-pass filter centered at $810 \pm 5$ nm.

Wavefront modulation is performed using a liquid-crystal-on-silicon spatial light modulator (Holoeye Pluto-NIR-II) featuring $1080 \times 1920$ pixels with an $8~\mu\text{m}$ pixel pitch. Single-photon detection is carried out with an SPC3 SPAD camera (Micro Photon Devices), comprising $32 \times 64$ pixels with a $150~\mu\text{m}$ pixel pitch. Intensity images are acquired using an Andor iXon Ultra EMCCD camera equipped with a $512 \times 512$-pixel sensor with a $16~\mu\text{m}$ pixel size.

The first lens illustrated in Figure~\ref{Figure1}.a, which Fourier-images the crystal onto the object, is in fact composed of a series of three lenses with focal lengths of $50$ mm, $150$ mm, and $100$ mm. The first two lenses are arranged in a $4f$ configuration, while the third lens images the Fourier plane of the crystal onto the object. The object is then Fourier-imaged onto the SLM using a $100$ mm lens. Finally, the last lens shown in Figure~\ref{Figure1}.a is also composed of three lenses with focal lengths of $200$ mm, $100$ mm, and $250$ mm. This combination provides a direct image of the object while preserving the Fourier imaging of the crystal plane.

\section*{Acknowledgments}

H.D. acknowledges funding from the ERC Starting Grant (No. SQIMIC-101039375) and the ANR (No. ANR-24-CE97-0001 and ANR-23-CE47-0014).

\bibliography{references}
 
\end{document}